# Developing and implementing an Einsteinian science curriculum from Years 3 to 10 – Part A: Concepts, rationale and learning outcomes

Tejinder Kaur[1], Magdalena Kersting[2], David Blair[1], Kyla Adams[1], David Treagust[3], Jesse Santoso[1], Anastasia Popkova[1], Shon Boublil[1], Marjan Zadnik[1], Li Ju[1], David Wood[1], Elaine Horne[1], Darren McGoran[1]

[1] School of Physics, University of Western Australia, Perth, Australia

[2] Department of Science Education, University of Copenhagen, Copenhagen, Denmark

[3] School of Education, Curtin University, Perth, Australia

**Abstract**

There has been a growing realisation that school science curricula do not adequately reflect the revolutionary changes in our scientific understanding of the 20th century. This discrepancy between current school education and our modern scientific understanding has led to calls for the modernisation of the science curriculum. Although there have been attempts to introduce topics of Einsteinian physics (i.e., quantum physics and relativity) to school education, often at the secondary level, we still lack a seamless curriculum in which modern science concepts are gradually introduced in primary and middle schools. Guided by the Model of Educational Reconstruction and following a mixed-methods research design, the Einstein-First project aims to address this gap. Einstein-First has developed and implemented an Einsteinian curriculum from Years 3 to 10 (students aged 7- 16) that resolves the disconnect between science in schools and the modern world. This paper presents the concepts, and rationale for the Einstein-First learning approach, as well as a summary of learning outcomes in six Australian schools with 315 students across Years 3 to 10. Our generally positive findings lay the foundation for informed curriculum development, and school education that provides all students with understanding and appreciation of the fundamental concepts that underpin the technologies of the modern world.

## 1. Introduction

### 1.1 The need for modernising the science curriculum

There is broad agreement about the need to modernise the science curriculum and educate a population that is scientifically literate population (Blandford & Thorne, 2020). One key reason is that current science curricula in primary and middle schools in Australia and many other nations predate the early 20th-century revolution in our scientific worldview and do not adequately reflect the current state of scientific knowledge (Kaur et al., 2018).

Einsteinian physics (EP), which encompasses quantum physics and the theory of relativity, has significantly shaped modern society and advanced technological progress. With its strong connection between fundamental principles and applications, EP is central to contemporary scientific disciplines, including quantum computing, materials science, biotechnology, multi-messenger astronomy, and climate science (Blandford & Thorne, 2020). Nevertheless, Einsteinian concepts are currently not taught to students beyond a superficial level. This discrepancy between current school education and our modern understanding can create confusion and roadblocks to science learning (Kersting & Blair, 2021).

Additionally, research has shown that students are motivated and excited to learn modern science concepts and are aware that the new concepts in the curriculum are meaningful to them (Choudhary et al., 2020; Kersting et al., 2018; Maxwell et al., 2021). Indeed, traditional science curricula often struggle to transmit a feeling of personal relevance and disciplinary authenticity (Kapon et al., 2018; Kaur et al., 2018; Kersting, Schrocker, et al., 2021), and student performance in science has decreased throughout the OECD nations in recent years (Kjærnsli & Lie, 2011; OECD, 2016). Furthermore, significant achievement gaps between students with different backgrounds and a decreasing number of students who choose science at the upper-secondary level give rise to concern (Treagust et al., 2015). It is widely suggested that this decline can be abated by introducing science that reflects public discourse, news items and modern technology (Foppoli et al., 2019; Sheldrake et al., 2017). A modernised science curriculum that includes Einsteinian Physics (hereafter abbreviated as EP) concepts can improve student attitudes and make science more engaging and relatable: not least because topics of EP can illustrate the nature of science and demonstrate the power and limitations of scientific reasoning and experimentation (Park et al., 2019; Woithe, Boselli, et al., 2022).

The widely recognised need to promote science diversity is another reason the school science curriculum needs to be modernised. Historically, females have been underrepresented in the natural sciences, and traditional curricula often reflect societal prejudices that presume science is the purview of men (Baram et al., 2011, Kaur et al., 2020; Ross & Gordon, 2018). Research has shown that topics of modern physics, which are of interest to girls and emphasise the relevance of science to everyday life, can increase the potential for students to identify with physics and help improve girls' attitudes and motivation towards science (Kaur et al., 2020; Kersting, Schrocker, et al., 2021; Woithe, Müller, et al., 2022; Zoechling et al., 2022). In summary, modernising the science curriculum is vital because it can align instruction with our current-best scientific understanding, improve student attitudes towards science, and promote gender equity in science education.

*1.2 Previous research on EP education*

Many studies have been conducted with upper-secondary and undergraduate students and have found that these students often struggle with the counterintuitive nature of EP (e.g., Ayene et al., 2019, Velentzas & Halkia, 2013). Students have difficulties connecting Einsteinian concepts to their knowledge of classical physics and everyday experiences of the world (e.g., Kamphorst et al., 2019; Krijtenburg-Lewerissa et al., 2017; Steier & Kersting, 2019). There is limited research on learning EP at an early age – when one would expect that students are not yet burdened with a prior scaffold of classical concepts (Kersting & Blair, 2021).

In 2011, the Einstein-First team trialled the first program with Year 6 students, and the results were emboldening: the students were not bewildered, but rather took the ideas in stride, indicating that EP can be taught to primary school students (Pitts et al., 2014, Adams et al., 2021, Blair, 2021). Other trials conducted with primary and middle-school students suggest that the teaching of EP concepts is largely independent of student aptitude or prior knowledge and has a lasting impact on the students' memory retention of EP (Choudhary et al., 2020; Haddad & Pella, 2015; Kaur et al., 2017b; Kaur et al., 2018; Pitts et al., 2014; Ruggiero et al., 2021, Adams et al., 2021). According to Metz, 'what students of any age are able to learn depends heavily on what they've already learned, failure to support the scientific capabilities of elementary school children will seriously handicap science learning at higher grade levels' (Metz, 2011, p. 71). Thus, the conceptual challenges that older students encounter may be due to having to switch ontological categories from a Newtonian to an Einsteinian conception of physical reality because of their classical school education (Kaur et al., 2018).

Overall, there is a growing body of literature on the teaching and learning of EP. However, there are still gaps, particularly in terms of understanding how students make connections between different EP concepts and how instruction can be designed to promote understanding of these concepts at an early age. Although there have been attempts to introduce specific topics of EP to school education, often at the secondary school level, we still lack a seamless curriculum in which modern science concepts build on each other meaningfully in primary and middle schools.

## 2. Research questions

This paper takes a significant step towards reconstructing science education from the ground up to introduce young learners to the contemporary Einsteinian paradigm that best describes our world and underpins contemporary scientific discourse (Treagust, 2021). We present the development, implementation, and evaluation of a seamless Einsteinian curriculum from Years 3 to 10 (students aged 7- 16 years) that resolves the disconnect between science in schools and the science of the modern world. This paper is structured in line with three research questions that unpack our educational reconstruction:

1. What instructional approaches and pedagogical principles are needed to effectively promote learning of EP at the primary and middle-school level?
2. What key science concepts and relationships should be emphasised in the EP curriculum to enable students from Years 3 to 10 to develop understanding of EP concepts?
3. To what extent do Year 3-10 students develop understanding of the main concepts of Einsteinian physics?

## 3. Educational background and theoretical framework

### 3.1 The Einstein-First project

This study is conducted within the Einstein-First project that brings together experts in physics, science education, curriculum development, and teacher training from three Australian universities in collaboration with several Australian teacher and school associations. Einstein-First aimed to trial and evaluate a progression of EP education through primary and middle school with research-informed teaching, learning materials and assessment instruments (Maxwell et al., 2021). By clarifying concepts, finding unifying principles, and using simple language to teach EP at an early age, Einstein-First replaces the pre-Einsteinian concepts implicit in primary school education with explicit introduction of Einsteinian concepts such as photons and the relativity of time. We introduce Einsteinian concepts and language in Years 3-6 (the last four years of Australian primary school) and revisit and develop them at an increasingly sophisticated level between Years 7 and 10 (the first four years of Australian secondary school). Year 10, the last year of compulsory science in Australia, is the last opportunity we have to ensure that the whole population can share a basic understanding of our best understanding of physical reality. At this point, we also want to encourage students' STEM subject choices.

### 3.2 The Model of Educational Reconstruction

The development of Einstein-First has been informed by the Model of Educational Reconstruction (MER), a widely used framework for science education research and development (Duit et al., 2012). Several studies in Einsteinian physics education have drawn on the model (e.g., Boublil et al. 2023, Kamphorst et al 2019, Kersting et al. 2018, Woithe et al., 2021). It is a good fit for our study since 'it has been developed as a theoretical framework for studies as to whether it is worthwhile and possible to teach particular content areas of science' (Duit et al., 2012, p. 19). The MER connects research on the science content structure to its educational importance, thereby building on empirical studies of student understanding and initial testing of pilot learning resources in authentic classroom settings. This interplay between research and development is reflected in the three strands of the model: 1) research on teaching and learning, 2) clarification of content, 3) evaluation of learning resources (Figure 1). These strands correspond to our three research questions which structure the following sections.

*Research Question 1: What Instructional approaches and pedagogical principles are needed to effectively promote learning of EP at the primary and middle-school level?*

In line with the MER, research on teaching and learning is key to identify instructional approaches and pedagogical principles that can promote learning of EP. We have adopted spiral learning as our overarching approach and synthesised existing research and our experiences from ten years of empirical trials in the form of pedagogical principles. Spiral learning is based on the idea that learning is a continuous process and that students should be exposed to new information at a developmentally appropriate level while revisiting and integrating previously learned material (Bruner, 1960; Harden, 1999). The move from simple to complex, from concrete to abstract, and from fundamental ideas to sound comprehension is a crucial benefit of spiral learning (Harden, 1999; Yumuşak, 2016), and one that aligns with the goal of the Einstein-First project. We wish to introduce students to the Einsteinian paradigm in a controlled way and at a level in which they can master the content before continuing to build new on prior knowledge.

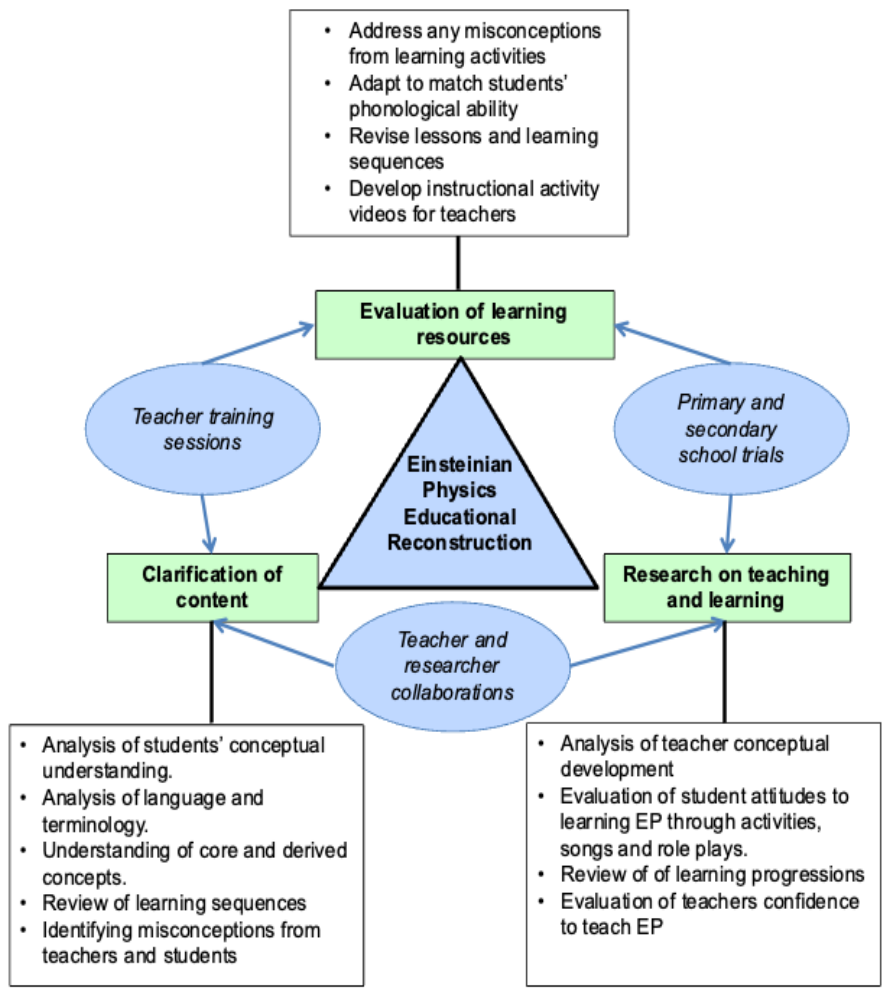

Figure 1: Our research design has followed the three components of the MER: clarification of content, research on teaching and learning, and development and evaluation of learning resources.

We acknowledge that the early years are crucial for language development, particularly in the context of modern science. It is imperative to gain scientific vocabulary during this time in order to foster the growth of scientific thinking (Lemke, 1990; Vygotsky, 1962). Thus, a crucial part of introducing Einsteinian physics in line with the spiral-learning approach is introducing Einsteinian language and, in places, selecting appropriate child-friendly terminology to describe Einsteinian phenomena. The following six pedagogical principles further underpin the development of our Einsteinian curriculum:

1. *Introduce concepts through activity-based group learning* We draw on an activity-led instructional approach that uses hands-on activities in small groups to introduce abstract concepts. Activity-based group learning is widely recognised to promote inclusivity, benefit female learners, and deepen conceptual understanding in Einsteinian physics (Alstein et al., 2021, Dare & Roehrig, 2016; Kaur et al., 2017b; Kersting et al., 2018, Labudde et al., 2000). Presenting concepts through activities makes it easy for students to understand the core concepts of Einsteinian physics. For example, our activities that introduce the concept of curved space consist of creating triangles on a curved surface (Image 1) and using toy car trajectories to map straight lines and see if parallel lines can cross (Image 2). Our students work in small groups, learn from each other, and generally report back in a teacher-led whole class discussion.

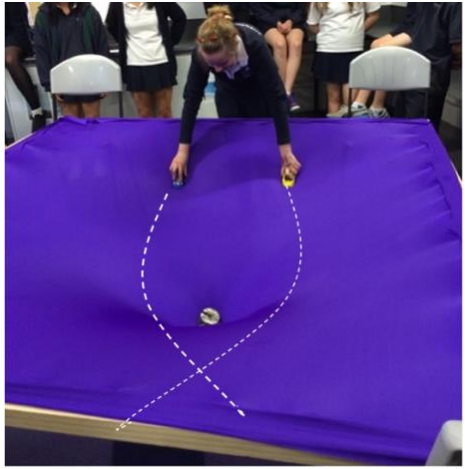

Image 1: Students learn curved space geometry by constructing triangles on a wok.

2. ***Use toys, models, and analogies*** All children understand toys and can distinguish easily between the toy and reality. For example, toy atoms, toy photons, and toy spacetime are appreciated by children and provide a powerful learning environment. The key to our approach is to foster intuitive insights by using children's innate ability to learn from toys and through play (Vygotsky, 2004). Besides, toys are in themselves models and analogies which have been shown to help students better understand abstract concepts (Aubusson et al., 2006). By providing simplified representations or comparisons to familiar ideas, models and analogies can become can become a powerful tool in Einsteinian physics education (García-Carmona, 2021; Kersting & Steier, 2018).

3. ***Promote whole-body learning*** When the activities involve whole-body learning, it provides students with tangible physical connections between scientific concepts and embodied experiences (Davis et al., 2019, Kersting et al., 2021, Rollinde, 2019). For example, at the Year 3 level, we introduce the concept of heat as jiggling molecules. We ask students to pretend to be molecules and to transfer their jiggling from person to person to enact thermal conduction. At the Year 8 level, we introduce the concepts of energy through activities involving the lifting of substantial masses to maximise the embodied experience where a thumping heart (or measured heart rate) indicates energy expenditure.

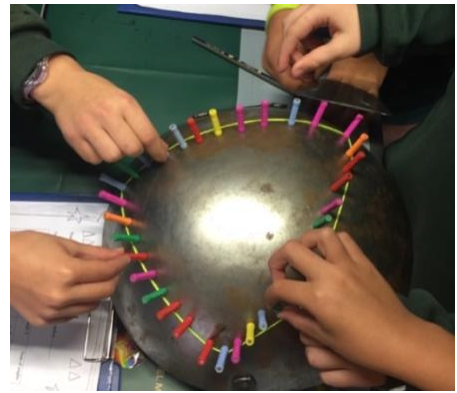

Image 2: Students map the trajectories of two pull back cars, which represent photons, on a curved surface.

4. ***Use appropriate language and keywords for unifying concepts*** We recognise that the historically derived language of Einsteinian physics is often opaque or inappropriate for young learners. By focusing on unifying principles and replacing traditional terminology with more descriptive terms with clearer meaning, we simplify the language of Einsteinian physics (Kersting & Blair, 2020). One unifying principle is the concept of binding energy which applies to interactions on all scales from nuclei to galaxies. We make binding energy tangible by using toy tennis-ball atoms containing powerful magnets that allow students to feel the binding, experience the work required to break the magnetic binding, and to play scattering games in which a toy photon can break apart a diatomic tennis-ball molecule. Another example for simplified terminology is the wave-particle duality that we introduce with an aphorism: *'Everything has waviness and everything has bulletiness'*. We make this idea concrete by having students work in groups to create 'pet photons' with wavy and bullety components (Image 3). These examples show how our pedagogical principles work together to provide students with engaging learning experiences.

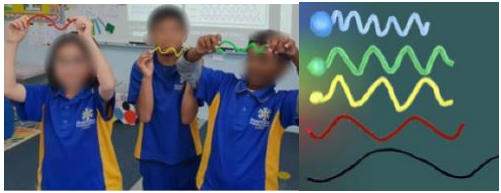

Image 3: Students representing the spectrum of photons from red to blue are learning about photons of different wavelengths. They are using pipe cleaners with plasticine masses to illustrate the varying amount of bulletiness.

5. ***Draw on role plays for representing conceptual change and human endeavour*** This principle relates to science as a story of conceptual changes and human endeavour over many centuries, involving successive struggles that have provided better and better understanding of physical reality. These stories are highly motivating for students (Kersting et al., 2018; Lin-Siegler et al., 2016) and can be emphasised using a dramatic device such as a time machine to bring the key historical figures together with questioning students. In a typical role play, students playing Euclid, Gauss, Riemann and Einstein debate with modern students, in a light-hearted context. Songs and chants make the plays memorable.

6. ***Use only inexpensive consumer-level equipment*** From balls and toy cars to solar panels and laser pointers, all apparatus in our program is simple, and all aspects of the program are easy to source and within school budgets. Using everyday materials makes the activities more relevant to primary schools, where specialised equipment is often not available (Palmer, 2006). From an educational point of view, doing science with everyday objects rather than highly specialized or expensive equipment has the added advantage of connecting science to students' lives (Ashbrook, 2003). Since many students perceive common school science to be disconnected from their everyday experiences, our principles aim to present science as being personally relevant and significant to the students (Kapon et al., 2018).

In summary, our pedagogical principles are powerful and effective because they connect abstract concepts to real-world experiences, uses toys, models, and analogies that students can relate to, promote both group learning and whole-body learning, simplify language where needed, and reduce the intellectual burden through the identification of unifying concepts. Inexpensive equipment is used for activity-led teacher education, making the program easily accessible, while role-plays engage students in the broad stories of paradigm change.

*Research Question 2. What key science concepts and relationships should be emphasised in the EP curriculum for students in Years 3 to 10 to learn EP concepts?*

To clarify the science content, (the second strand of the MER), a panel of physicists and physics educators determined the initial content sequencing for teaching in line with the spiral curriculum approach. Our challenge was to find appropriate ways to identify the core Einsteinian concepts and modernise the school curriculum without drastic reformulation. Therefore, we identified the main thematic content and enhanced each theme in the context of modern science relevant to current scientific discourse. Crucial to this modernisation was the identification of the following: a) invalid and obsolete concepts, b) key concepts and subject areas essential for understanding modern science and technology, c) unifying concepts that can contribute to understanding a broad range of science, d) appropriate models and toys that can enable activity-based learning, e) progressions of learning that allow spiral learning to reinforce understanding over eight years of science education.

In Figure 2 we summarise the Einstein-First curriculum in a simple spiral learning format. The core concepts such as light being a stream of photons, are introduced first, and repeated and developed and combined into more advanced concepts. Figure 3 presents the curriculum in more detail, showing links between concepts as well as associated mathematics. Both figures illustrate the progression from core concepts to derived concepts. See the figure caption for more detail.

In parallel with clarifying the content, we developed lesson plans in which the central focus is on the activities, with the concepts being revealed by the activity. The activities, toys, and models included in the lessons were developed and tested through two or three iterations of the MER process. Teachers (primary or secondary) attended informal training sessions, or participated in trial lessons, to identify those activities that were most or least effective in revealing the relevant concepts (Boublil & Blair, 2023; Choudhary & Blair, 2021; Kaur et al., 2017). The modules were refined based on the teacher feedback and revised again after formal school trial implementation[1].

The spiral learning approach allows for introduction, development and reinforcement of Einsteinian concepts. By revisiting and building on concepts at different grade levels, students are given multiple opportunities to understand and internalise them.

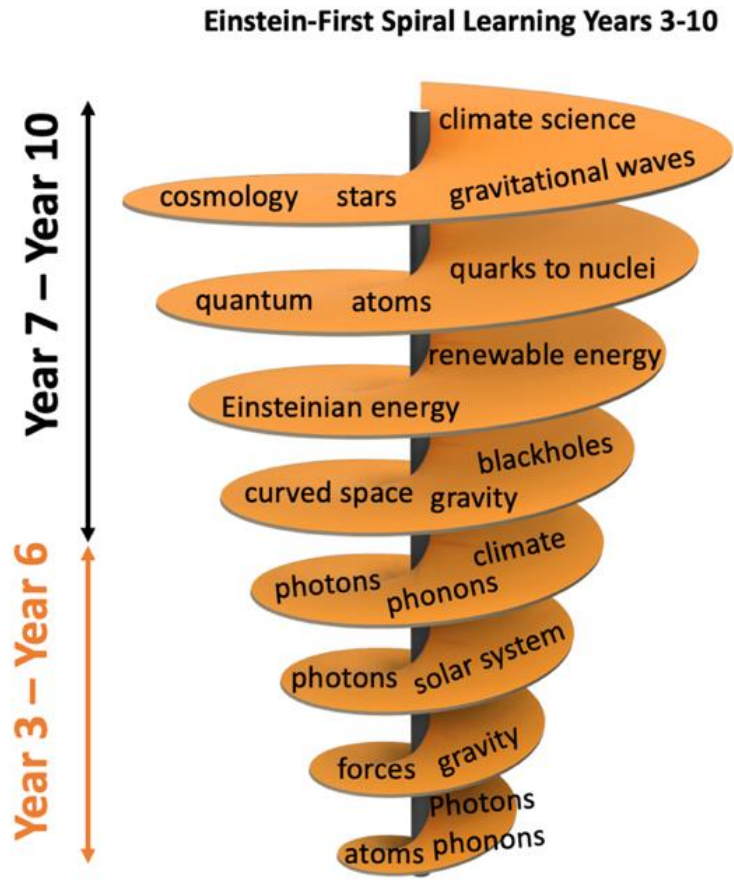

Figure 2: A spiral diagram adapted for Einsteinian curriculum in years 3 – 10. Concepts are introduced through toys and Einsteinian language in years 3-6 and then revisited at an increasingly sophisticated level between years 7 and 10. For example, phonons and photons are introduced in year 3, and revisited in years 5, 6, and 9.

---

[1] The lesson plans and assessment tools are accessible on the Einstein-First website within a password protected area, which can be accessed by sending an email to the first author.

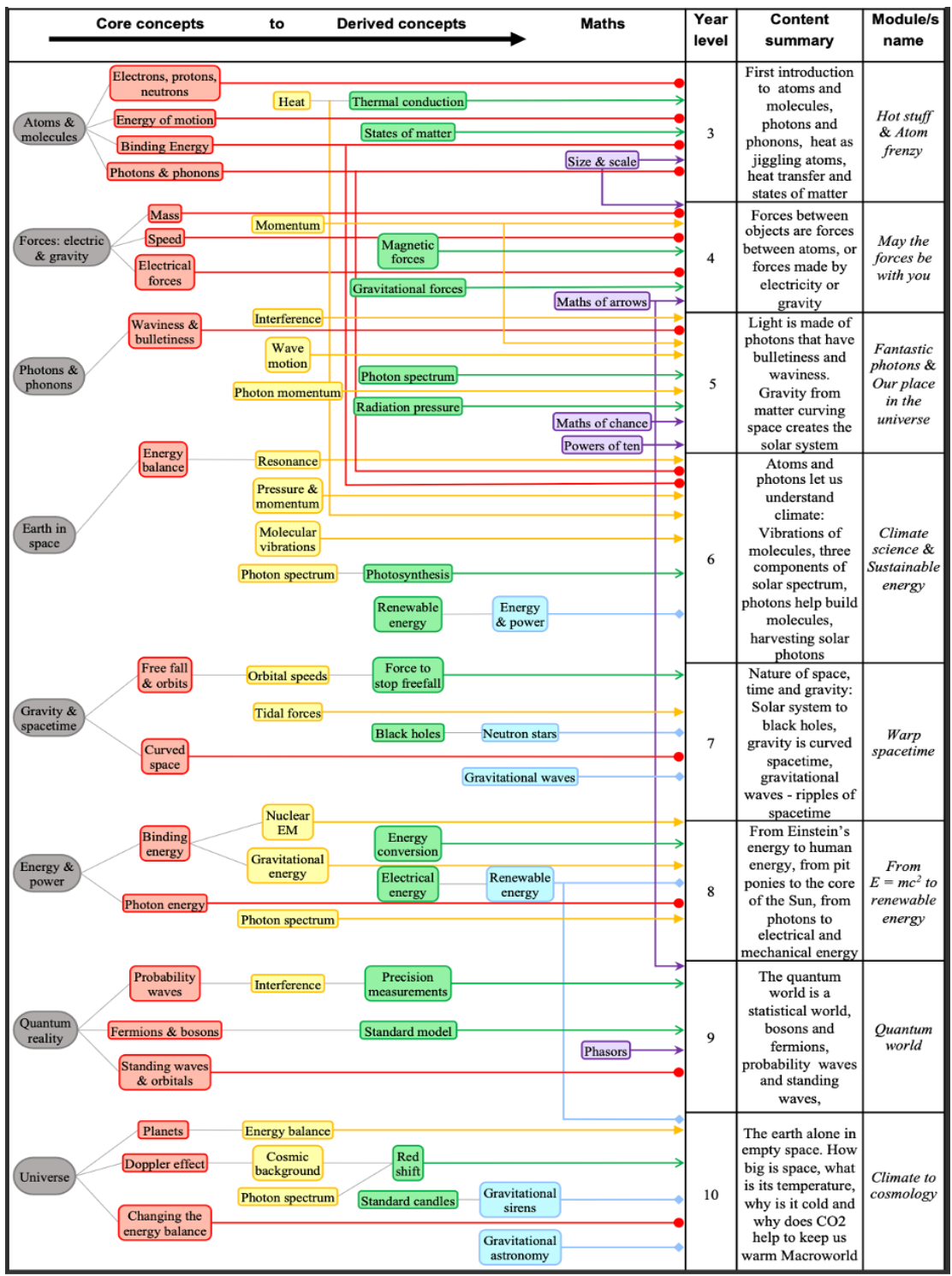

Figure 3: An outline of the 8-year Einstein-First program from Year 3 to Year 10. On the far left is a list of themes and concepts. The red boxes display core concepts with yellow, green to blue, indicating a progression to derived concepts. Connectors and repetitions show how the concepts introduced in early years are developed in later years. The pale purple boxes indicate where associated mathematical concepts ( simple vectors called the maths of arrows, probability and powers of ten) are introduced and then developed in future years according to the pale purple arrows.

## 4. Methods

To evaluate our modules empirically, we trialled the lesson plans in primary and middle schools (Table 2). Trials for Years 3-5 were presented by teachers who had recently trained in 1–2-day workshops, while Einstein-First personnel were available to advise the teachers who presented the trials for years 7, 8 and 10 programs. The Year-9 program was delivered by one of the researchers, with classroom science teachers in attendance during the trials. The trials varied in length from eight to ten lessons. Each primary module had eight lessons delivered over a term: one 60-minute lesson every week, and secondary school modules (except the Year 9) had 10-12 lessons spread out over a term, with two 60-minute lessons per week on average. The Year 9 trial consisted of eight lessons with one 45-minute lesson per week. All trials used a standard set of interactive activities designed for each year group, similar lesson plans[2] and a consistent pre- and post-test procedure for each year group. After each trial, the tests were revised based on teacher feedback and test results. This procedure is consistent with the MER framework: the revisions identified unintended ambiguities in the wording but did not change items in such a way to invalidate the concept being measured.

### 4.1 Data collection

Each trial consisted of three phases of data collection. The first phase consisted of pre-test instruments. These instruments were used to obtain a baseline understanding of student knowledge of and attitudes. The second phase consisted of post-test instruments. These provided a measure of the student knowledge and attitudes at the end of the trial period. Both phase 1 and phase 2 data was conducted in the usual classroom context. The research team then marked the knowledge tests using a pre-determined and evaluated marking guide. The third phase consisted of teacher feedback. This took the form of formal interviews of the teachers who presented or participated in the trials and their written feedback on sheets at the end of professional development sessions. Teacher interviews were conducted using questions designed to probe the interviewee's confidence in teaching EP and gain insight into their experience of the teaching and learning approaches used at the different year levels. These interviews helped shape alterations to the lesson content and guided the focus of the analysis. A detailed analysis of teachers' experiences is presented in the accompanying paper, which provides more in-depth information (Kaur et al, submitted).

Table 2 Trial information, showing the year levels, module names, and basic data for the lessons analysed in this paper.

| Year level | Modules reported in this paper | Number of lessons | Student sample | Number of classes | Teacher Interview |
|---|---|---|---|---|---|
| 3 | Hot stuff | 8 | 57 | 3 | Yes |
| 4 | May the forces be with you | 8 | 66 | 3 | Yes |
| 5 | Fantastic photons | 8 | 30 | 1 | Yes |
| 7 | Warp spacetime | 12 | 80 | 3 | Yes |
| 8 | From E = $mc^2$ to renewable energy | 10 | 22 | 1 | Yes |
| 9 | Quantum world | 8 | 42 | 3 | Yes |
| 10 | Cosmology | 8 | 18 | 1 | No |

### 4.2 Pre- and post-tests

Pre- and post-knowledge tests were designed to provide a measure of student understanding of relevant core and derived concepts. The tests consisted of open-ended and multiple-choice questions. The language of the questions was chosen based on the year level and to ensure that understanding of the desired concept was measured. The majority of the questions in the pre- and post-tests were the same; however, the post-test often had extra questions

---

[2] Both primary and secondary school lessons rely on hands-on activities. The lesson plans suggest two or more activities, and the teacher may implement one or all of them depending on the available class time.

The team created PowerPoint slides for the teachers to introduce the topic, and lessons included worksheets that students can complete during or after the activities.

as a result of introducing concepts to year levels where there could reasonably be no baseline measure expected. Attitudinal tests were also administered but will not be reported here as they are outside the scope of this paper.

## 4.3 Data analysis

Pre- and post- knowledge tests were administered by the teachers but marked by members of the research team. The team recorded students' responses and scores on an excel spreadsheet. The mark guide typically had the following structure:

- No mark for a blank or incorrect response, or not using Einsteinian language.
- Half marks for a partially correct and Einsteinian-language-consistent response.
- Full marks for a correct and Einsteinian-language-consistent response.

Once the tests were marked, the total score was converted to a percentage to account for the different question types and numbers between pre- and post-test, and to make relative comparisons across year levels manageable. Student data was anonymised. The student scores for each module were arranged by increasing pre-test score. The post-test score for the same student is then reported next to the relevant pre-test score. The data presented is a combination of trials for each module (i.e., not just a single class or teacher trial).

The procedure of direct comparison between ranked pre-test and post-tests provides a powerful means of rapid assessment of lesson outcomes. The shape of the pre-test histogram measures the newness of the content (low scores) and the level of outside-school learning (scores at the high end). Because we aim for learning by *all* students rather than trying to discriminate students according to ability, we are encouraged when the post-test scores are high and uncorrelated with pre-test score. This informs, without the need for statistical analysis whether we are meeting the goal of making Einsteinian concepts available to all students. The ranked pre-test/post-test comparison is also useful in understanding the school demographic, because pre-test histograms with larger high-score shoulders tend to correlate with higher SES demographics.

By studying individual question scores, it is possible to identify problems, sometimes due to insufficient teacher in-service education, and sometimes due to lack of emphasis in lesson plans. These observations enable appropriate modification of the lesson plans and teacher in-service education. It is beyond the scope of this paper to discuss results at the individual question level, but in the discussion below we will give examples of individual questions that were used to identify areas of difficulty. In each case, the results were used to improve lesson plans, clarify a teacher's misconceptions and in a few instances to improve the wording of tests.

*Research Question 3: To what extent do Year 3-10 students develop understanding of the main concepts of Einsteinian physics?*

We will now examine ranked pre-test/post test data as described above, for trials across 7 of the 8 years of the Einstein-First curriculum. Data from each trial set identified in Table 2 is presented in Figures 4a – 4g. An analysis of the test outcomes for questions relating to the identified main concepts presented in Figure 3 allows us to consider the overall outcomes of the Einstein-First teaching approach. We will show how it was used to detect issues that caused confusion or limited learning, thereby allowing improvements of the instructional approaches.

Figures 4 a-g show pre and post-test scores of Years 3 – 5 and Years 7 – 10 students in ascending order of their pre-test score. Each year level result is broken down below into a few key concepts from that module. Selected relevant questions and the student test score outcomes are discussed to highlight the changes in student scores after the instruction of EP. Detailed analyses of student learning are left the subject of future work.

*Year 3 (age 8):* Figure 4a shows results for three classes of Year 3 students who were taught the "Hot stuff" module. In response to the question "Describe what you think an atom is?" students gave answers such as, "very tiny objects", "invisible things that make everything". More than 70% of students were able to describe atoms. In response to the question "Describe what you think a molecule is?", approximately, 50% of the participants could describe a molecule. Some key phrases were: "group of atoms", "molecules are bigger than atoms". During our testing, we identified phonological difficulties with the words, photon, phonon, and proton, which are discussed under Identification of Problems below. In response to the difficulties, we created new songs to help students disentangle the similar words with very different meanings.

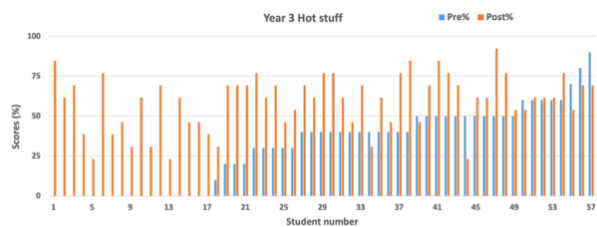
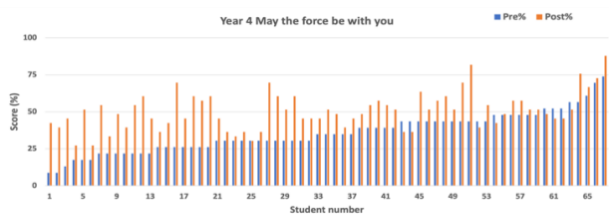

Figures 4(a) and 4(b) present the results of Year 3 and Year 4 students before and after the program.

*Year 4 (age 9)*: Figure 4b shows results for three classes of Year 4 students who were taught the "May the forces be with you" module. The histogram is somewhat anomalous compared with other years. There was a high level of pre-knowledge, probably because students at this age often have informally learnt knowledge of electrical forces and gravity. Students with low pre-knowledge showed substantial improvement, but students with high pre-knowledge showed negligible improvement.

The results were used to identify overly complex test questions and misconceptions caused by confusion between traditional learning about push-pull forces and our Einsteinian activities about gravity and electrical forces. This led to substantial revisions of the teacher training and the course content.

One of the questions assessed students' understanding of distance, speed, time, acceleration, inertia, and friction. Students were given multiple answer options and instructed to select the correct response for each statement. After the program, nearly 98% of students demonstrated comprehension of distance, speed, and time concepts, and 78% of students understood acceleration. For the remaining two statements, students were aware that the answers would be either friction or inertia but appeared confused as to which option to select. Only 55% of students were able to comprehend the meanings of inertia and friction.

*Year 5 (age 10)*: Figure 4c shows results for one class of Year 5 students who were taught the "Fantastic photons" module. After participating in the program, every student in the class was able to correctly identify photons as the correct response to the question "A little over 100 years ago Einstein told us that light comes as little particles. What are they called"? In response to the question, "What do you think is the fastest thing in this universe?", all the students cited light. A core concept for understanding the photoelectric effect is the fact that photons have momentum and by colliding with electrons, they transfer energy to them. Several class activities focussed on this. A key to teaching this concept is an association of the word momentum and its effect: the transfer of energy from one thing to another. One question "Can light push things? Please explain your answer" was asked to assess students' understanding of momentum. We were surprised that only 23% of students were able to explain that light is made up of photons and that they have momentum, allowing light to push things, because previous studies with interventions led by members of the project team much higher results were obtained (Choudhary et al, 2020). In this case, the low score on momentum was used to improve the lesson plans to make a stronger connection between the word momentum and an activity involving momentum transfer where insufficient emphasise on language was identified as the cause of the low score.

*Year 7(age 12)*: Figure 4d shows results of 80 students of Year 7 (3 classes) who were taught the "Warp spacetime - Gravity from the Earth to black holes" module. This program mostly explored the concepts of curved geometry, free-fall, motion, and curved spacetime. Teachers did not have enough time to explore all the activities using the spacetime simulator and the activities on gravitational waves. Nevertheless, we observed a large improvement in students individual scores after the program, indicating that the program positively affected all students irrespective of their pre-knowledge or ability. In this program we discovered two areas of confusion that had not been apparent in previous teaching of the same material. We learnt from the comparison of previous programs (Kaur et al 2018) that it is particularly important to emphasise the break between the Newtonian/Euclidean paradigm and the Einsteinian paradigm.

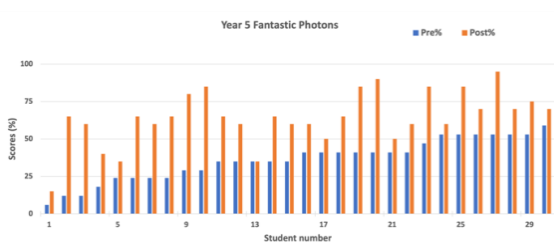
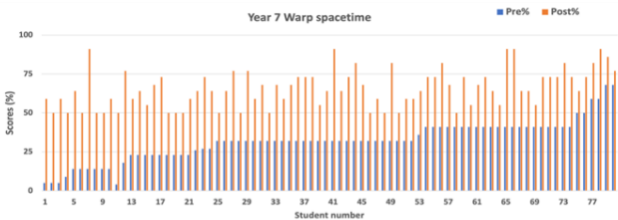

Figures 4(c) and 4(d) present the results of Year 5 and Year 7 students before and after the program.

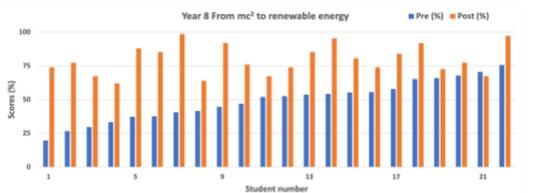
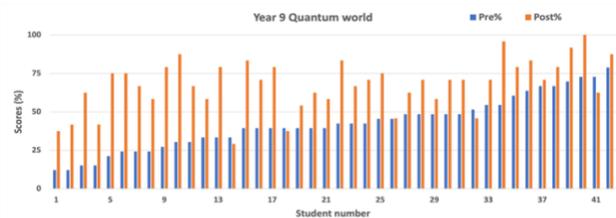

Figures 4(e) and 4(f) present the results of year 5 and year 7 students before and after the program.

In this program, the teacher in-service material failed to emphasise the measured and demonstrated invalidity of Newtonian gravity, leaving students clearly confused, as demonstrated by their answers to questions about gravity and geometry. Their concurrent mathematics program on Euclidean geometry conflicted with the curved space geometry taught in this program, accentuating the confusion. Further elaboration on these confusions can be found in the following section.

*Year 8 (age 13)*: Figure 4e shows results for one class of Year 8 students who were taught the "Einsteinian Energy" module. The Year 8 program begins by confronting students with the two equations of Einsteinian energy, $E=mc^2$ and $E=hf$. This is students first exposure to equations in physics, and although only equations of proportionality, they involve huge and tiny constants. The equations and their units are connected to the historical discovery of energy concepts by James Watt. The program goes on to cover renewable energy, photon physics and binding energy.

The results demonstrate that students have a moderate understanding of Einsteinian energy topics prior to the program and make substantial progress after the series of lessons. There were a few maths questions asked to the students. One of them was based on the equation $E=mc^2$, and the other on the $E=hf$. Students were given the sun's power output and asked to calculate the mass lost by the sun using the equation $E=mc^2$. Only 43% of students solved the problem after the program. In another question, students were asked to calculate the energy in joules of an ultraviolet photon of frequency $10^{16}$ Hertz using $E = hf$, where Planck's constant $h = 7 \times 10^{-34} \text{JHz}^{-1}$. After the program, 77% were able to solve it correctly.

*Year 9 (age 14)*: Figure 4f presents the pre- and post-test results for three Year 9 classes from two different schools who were taught the "Quantum world" module. The focus of the module was the quantum description of light. The results of the study indicate that students were generally able to use appropriate EP language when describing photons as quantum objects with both wave and particle characteristics has improved. In response to the question "What are photons?" 50% of students demonstrated their comprehension by providing answers consistent with EP, whereas 33% of responses were partially consistent. 74% of students were able to define waves, and 76% of students were able to explain the relationship between photons and waves using EP language. To assess students' EP learning relating to solar panels, the question "How do solar panels generate electricity?" was asked. Nearly 50% of students were able to describe how solar panels generate electricity.

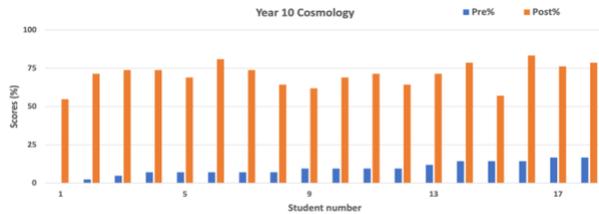

Figures 4(g) presents the results of year 10 students before and after the program.

*Year 10 (age 15)*: Figure 4g shows results for one class of Year 10 students' results who were taught cosmological concepts to give students the foundations to understand gravitational waves and how they can be used to determine the Hubble Constant. They examined how Hubble's Constant can be used to give a clearer understanding of the beginning and future of our universe. The data shows that the students had very little knowledge of cosmology before beginning the program. They were able to make significant gains in their understanding of the concepts covered in the lessons.

## 5. Discussion

In the previous section, we described how we learnt from trials across 15 classes and 7 year levels of Einstein-First. All the trials showed substantial gains in knowledge, supporting the case that classroom teachers can be upskilled sufficiently to effectively teach Einsteinian physics, which in all cases had not been part of their own education.

However analysis of the program rand individual post-test results revealed a number of issues which we discuss below. In all cases they have been used to improve lesson plans, mainly through changes of emphasis.

1. Avoiding bias from diverse reading skill: at the Year 3 and 4 level, there is high diversity in student reading levels. Tests results were clearly influenced by student reading difficulties, leading to revised tests that were more pictorial to minimise dependence on reading skills.

2. Year 3 students typically have a limited vocabulary and little phonological awareness (Townend 2000). This was reflected in their answers involving the words "photons", "phonons", and "protons". In response we created songs to help students disentangle the similar words with very different meanings[3].

3. Because inertia and friction are both associated with resistance to motion, we found that some students in year 4 and 5 struggled to distinguish between them. It is necessary that activities clearly separate and distinguish large mass low friction motion such as a child's motion on a playground swing, from low mass high friction situations such as dragging a chair across a carpet. These ideas then link to ideas that inertia is a coupling between mass and space, while friction is about interatomic forces.

4. More attention needs to be given to dimensionality and the essential differences between 3D and 2D spaces in teaching Einsteinian gravity. It is easy to assume students intuitively recognise the differences between 2D and 3D spaces, but this idea needs to be carefully unpacked when we connect trajectories on 2D surfaces with trajectories in 3D space.

5. Language problems also occur due to assumed meaning of the terms "law" and "theory". The terms Kepler's laws (eg. that planets move in ellipses) and Newton's universal law of gravitation carry with them an implication of correctness, even though both are violated in Einsteinian gravity and the real world, because space is curved. Terms such as Einstein's general *theory* of relativity, imply tentativeness and conjecture. Historically this interpretation was true, but today general relativity is one of the best tested theories in physics and clearly supercedes Newtonian gravity. Despite the fact that Newtonian approximations have utility, it is important that the radical differences be emphssised: space is like an elastic fabric rather than some sort of absolute framework. While our activities using lycra sheet models present the modern concepts, difficulties still arise as discussed below.

---

[3] https://www.youtube.com/watch?v=AxSgElLe9Ig

6. The above confusion was apparent in Year 7 students' results, because Year 7 Einsteinian gravity contradicted simultaneous learning of Euclidean geometry in mathematics lessons. Lycra sheet models using parallel trajectories of toy cars clearly demonstrate gravitational lensing and violation of Euclid's fifth postulate. Images of gravitational lensing, and digital learning resources reinforce the reality of this phenomenon (Kaur et al. 2017, Kersting et al. 2018). Yet in post-tests, many students wrote that parallel lines can never intersect, with some explicity stating that that was what the mathematics teacher told them. It is evident that there must be a stronger emphasis on the differences between theoretical geometry on flat paper and the geometry of real space observed by astronomers.

7. We also discovered that the concept of gravity could be easily confused if there was not very careful and consistent messaging. Students in the Year 4 trials were easily able to comprehend how lycra sheet models becomes curved when mass is added, but many were unable to explain what this activity tells us about gravity. In addition, we discovered that Year 7 students also struggled to explain gravity as curvature of spacetime. Our findings are consistent with previous work in Norway with high school physics students, in which educators reported that spacetime simulator activities are problematic due to students' reliance on classical gravity to comprehend a new interpretation of gravity (Kersting et al. 2018).

The results reported here are designed to support the overview of Einstein-First presented in this paper. The test results presented here do not capture the depth of conceptual understanding or the ability to apply the learned concepts. Forthcoming papers will present more detailed analysis of each module, including further data based on teacher and student interviews.

Our results do not measure the long-term impact and sustainability of the intervention on students' learning. However after the original researcher-led trials we did conducted three year and ten year evaluations of student learning that gave positive results (Kaur et al. 2018; Adams et al. 2021). We plan a longitudinal study to measure changes in teenage student aspirations in relation to STEM studies and future careers.

This study relied on data from teacher-delivered programs. This can introduce potential biases in the data collected. For example, we had an instance where a teacher prepared students in advance for the pre-test. This was quite obvious in the anomalous pre-test histogram and results from this class were excluded from this study. Despite our teacher training, variations in their instructional skills, teaching styles, and implementation fidelity clearly influence classroom outcomes. However the overall consistency of the results across several schools indicates the general effectiveness of teaching Einsteinian physics using hands-on activities.

### 6. Conclusion

We have presented an overview of an eight year curriculum designed to replace implicit learning of pre-Einsteinian concepts with a curriculum designed to be overtly and uniformly Einsteinian, beginning with atoms, molecules, photons, phonons, curved space and warped time and ending with programs teach concepts from black holes to quantum entanglement, cosmology to quantum technologies. The curriculum uses the principles of spiral learning, in which key concepts and instructional activities are gradually introduced and built upon over eight years. It is a response to the urgent need for fostering science-literate students based on evidence that introducing young learners to the contemporary Einsteinian paradigm provides an excellent way of boosting student enthusiasm for science (Kaur et al. 2018).

To develop the Einsteinian curriculum, we followed Treagust's advice that for introducing "a new school curriculum such as Einsteinian physics […] it is important to take a broad perspective and examine new theories and ideas about teaching and learning that have been introduced in education and the ways to measure the success or failure of this introduction using educational research' (Treagust, 2021, p. 17). This advice was implemented in the form of six pedagogical principles that informed the entire program development, while the MER framework was used with trial results and teacher training: classroom results identified weaknesses, and these in turn enabled another cycle of trials using improved lesson plans and teacher training.

A crucial part of this study was the in-service education of teachers and their responses to the program which is not emphasised in this paper. An accompanying paper B analyses the teacher responses.(Kaur et al.,) and gives further insights into student learning through the observations of the classroom teachers.

Trial results across a range of classes and school years provided evidence of the effectiveness of our Einsteinian curriculum, while uncovering many areas for improvement. They support the idea that modern physics concepts can be meaningfully included in science education from an early age.

It is important to note that our results do not test the full spiral curriculum because there has been insufficient time to test the 8-year learning progression proposed. Spiral learning over the full 8-years should be more effective than the single year trials presented here. We did not present year 6 results because this program depends more strongly on the preceeding years.

In general, students demonstrated rapid acceptance of the Einsteinian vocabulary, except for areas of phonological confusion we discussed, addressed by introducing songs designed to separate similar words.

To our knowledge, this paper represents the first holistic approach to a consistent introduction of Einsteinian physics to replace the current pre-Einsteinian curricula mainly developed more than a century ago. Previous research suggested educational reconstructions of individual Einsteinian physics topics for secondary-school students (e.g., Boublil et al., 2023; Kamphorst et al., 2021; Kersting et al., 2018; Laherto, 2010; Woithe & Kersting, 2021). This paper presented a coherent and integrated curriculum consistent with modern scientific knowledge in a relatively seamless, logical progression of ideas that build upon one another from years 3 to 10.

Future work in this area will focus on expanding the implementation of the Einsteinian curriculum in more schools and gathering more data on the long-term impacts of the curriculum on students' learning outcomes.

The implications of this study for science education are clear: it is vital and possible to modernise science curricula to include the most recent scientific concepts and theories. Our primary motivation is to modernise science literacy so that everyone can engage with the wonderful revelations of modern science. A curriculum that teaches the fundamental principles of our current scientific worldview can become the educational engine for future innovation and discovery. Further, longer-term research will be needed to understand how this curriculum affects students' attitudes towards physics, and particularly their future career choices.

## Acknowledgments

We would like to thank the Einstein-First team, the ARC (2019-2024) for funding this work, our ARC Centre of Excellence for Gravitational-Wave Discovery (OzGrav) colleagues as well as our EPER collaborator for their enthusiastic support. We are grateful to our participating schools, principals, teachers, and students for allowing us to conduct the program and for granting us permission to use their photographs and data for research purposes. Our program is supported by numerous companies listed on our website www. einsteinianphysics.com.

## Ethics statement

All of the participating schools' principals, teachers, students, and parents provided their informed consent. In addition, students were informed that participation in the Einstein-First project was voluntary and would not affect their school grades. The research design followed ethical standards-compliant with the Research Ethics Committee at the University of Western Australia (2019/RA/4/20/5875), the Department of Education Western Australia, Association of Independent schools of Western Australia and the Catholic Education Western Australia. All the data are stored securely on the Institutional Research Data Store (IRDS) system at the University of Western Australia.